\documentclass[%
floatfix,
aps,
jor,
amsmath,amssymb,
showkeys
]{revtex4-2} 

\usepackage{amsmath,amssymb}
\usepackage{graphicx}
\usepackage{dcolumn}
\usepackage{bm}
\usepackage{amsmath}
\usepackage{subfig}
\usepackage{floatrow}
\usepackage{hyperref}
\usepackage[autostyle]{csquotes}
\usepackage{microtype}
\usepackage{comment}
\usepackage{gensymb}
\usepackage{tikz}
\usepackage{subcaption}

\hypersetup{
    colorlinks=true,
    citecolor=red,
    linkcolor=blue,
    filecolor=magenta,      
    urlcolor=cyan,
    pdftitle={Overleaf Example},
    pdfpagemode=FullScreen,
    }

\begin{document}

\preprint{AIP/123-QED}

\title[Sample title]{\textbf{A model of thermophoresis of colloidal proteins in water using non-Fickian diffusion currents}
}

\author{Mayank Sharma}
 \email{mayank.sharma@students.iiserpune.ac.in}
  \author{Angad Singh}%
  \email{angad.singh@students.iiserpune.ac.in}
 \author{A. Bhattacharyay}%
  \email{a.bhattacharyay@iiserpune.ac.in}
 \affiliation{ 
 Indian Institute of Science Education and Research, Pune, India
 }%

\date{\today}

\begin{abstract}
 In 1928, Chapman generalised Einstein's theory of diffusion for non-uniform fluids to show the presence of a non-Fickian diffusion current, which he considered important in thermodiffusion (Ludwig-Soret effect). In 1941, Kiyosi It\^o proposed the formal methods of stochastic calculus in the presence of spatially dependent diffusion, yielding the same non-Fickian diffusion current as shown by Chapman. The phenomena of thermodiffusion and thermophoresis occur in the presence of a temperature gradient, making diffusion space-dependent. The role of solvation forces in thermophoresis generated at the particle-liquid interface can be clearly understood once diffusion is properly understood. In this paper, we investigate the importance of Chapman's non-Fickian diffusion current for the thermophoretic motion of colloidal particles in water (at low salt concentration). We show that all the general features of variations of the Soret coefficient $S_T$ with temperature can be captured using Chapman's non-Fickian diffusion current. We compare our theoretical results with experimental plots of the Soret coefficients for three polypeptides in aqueous solution: Lysozyme, BLGA, and Poly-L-Lysine, and find a strong match. We emphasise that, in addition to the as-yet-understood details of solvation forces, Chapman's non-Fickian diffusion current is an indispensable element that must be taken into account for a complete understanding of thermophoresis. 

\end{abstract}

\keywords{Thermophoresis, Coordinate-dependent damping,  Soret coefficient, Colloidal proteins, Brownian Motion}
\maketitle

\section*{INTRODUCTION}

\begin{figure*}[t]
\centering
\begin{minipage}{0.8\textwidth}
    \centering
    \includegraphics[width=\textwidth]{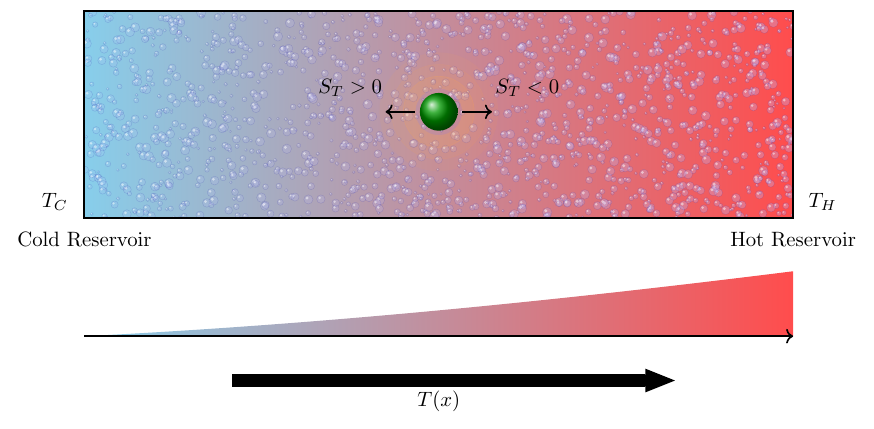}\\[0.25cm]
\end{minipage}

\caption{
    Schematic illustration of thermophoresis.
}
\label{fig:three_schematics}
\end{figure*}

Thermophoretic motion of colloidal particles in a solvent is a complex phenomenon that has a sensitive dependence on particles' diffusivity in the presence of a temperature gradient in the fluid, along with the physics in the solvation layer of the fluid that forms around the surface of the colloidal particles \cite{zheng2002thermophoresis,piazza2008thermophoresis,piazza2004thermal,piazza2008thermophoresis2,rahman2014thermodiffusion, niether2019thermophoresis}. There could be myriad mechanisms in the fluid's solvation layer around a particle, involving geometry, chemistry, charge distribution, electrostatic forces, locally induced hydrodynamics, etc., in the presence of a temperature gradient. This widens the scope of controlling particle transport by adjusting several parameters in a fluid under the general presence of a temperature gradient, making the understanding of the process essential for numerous applications and engineering \cite{vigolo2010thermophoresis,wienken2010protein,jerabek2011molecular,asmari2018thermophoresis,amaya,nehme5448133microscale,li2026thermophoretic}. The phenomenon of thermophoresis plays a significant role across a wide range of systems, from biological processes such as protein transport, biomolecular interactions, and cellular organization to technological applications including microfluidic separation, drug delivery, and materials processing\cite{zheng2002thermophoresis,asmari2018thermophoresis, franzl2022hydrodynamic, lin2016light, pavan2024brownian, chand2023emergence, magnez2022microscale, shakib2021microscale, jerabek2011molecular, niether2019thermophoresis, magnez2022microscale}. Its sensitivity to particle properties and solvent conditions makes it a powerful tool for manipulating and characterising microscopic entities. Despite extensive studies, the underlying microscopic mechanisms governing thermophoresis remain not fully understood, making it an active area of research with both fundamental and practical importance. There exist theoretical works using the application of kinetic theory \cite{bringuier2003colloid,bringuier2007kinetic}, continuum approach \cite{BRENNER2005251}, physico-chemical mechanics \cite{10.1063/5.0028674}, a framework on the basis of Stefan-Maxwell equation \cite{https://doi.org/10.1002/aic.17599}, and thermodynamic theory \cite{Shapiro+2020+343+372}. This is a typical non-equilibrium phenomenon thought to involve the transport of colloidal particles, largely due to the interplay between competing Fickian diffusion and solvation forces. Generally, experiments track the steady-state local particle density ($C(T)$) in the solvent at temperature ($T$). The central parameter, the Soret coefficient $S_T(T) = -1/C(T)[dC(T)/dT]$, quantifies the phenomenon of thermophoresis. The Soret coefficient, as a function of temperature for thermophoresis, quantifies how strongly the temperature gradient alters the concentration homogeneity that would be present in its absence. Figure 1, schematically represents the thermophilic/thermophobic situations where the phoretic particle moves along/opposite to the applied temperature gradient in the solvent.

The thermophoretic motion of colloidal particles is closely related to thermodiffusion, first reported by Ludwig in 1859 \cite{ludwig1856difusion} and subsequently observed by Soret in 1879 \cite{soret1879etat}. The process of thermodiffusion, also known as the Ludwig-Soret effect, involves partial segregation of components in a solution or fluid mixture in the presence of a thermal gradient. Thermodiffusion is understood to play a significant role in the convective separation of components in a fluid mixture. It is the interplay of convective flow and diffusion that underlie the thermodiffusive partial separation of the mixture's components. Therefore, a proper theoretical understanding of these complex processes of thermodiffusion and thermophoresis could only have meaningfully begun not before the advent of the physical theory of diffusion based on Einstein's work on Brownian motion in 1905 \cite{einstein1905motion, einstein1956investigations}. As is well acknowledged, Einstein's theory of Brownian motion provides the microscopic origin of Fickian diffusion. What lacks wider acknowledgement is that Einstein's theory of Brownian motion also underlies the non-Fickian diffusion in inhomogeneous media and fields.

Significant early work on the theory of thermodiffusion was undertaken by Sydney Chapman \cite{10.1098/rsta.1912.0012} using kinetic theory. In 1928, Chapman published a work in which he generalised Einstein's theory of diffusion for particles in a non-uniform field \cite{chapman1928brownian}. Chapman had generalised Einstein's theory of diffusion, with particular emphasis on its suitability for thermodiffusion problems. In this paper, Chapman clearly showed, based on kinetic theory, the presence of a non-Fickian diffusion current density proportional to the diffusivity gradient. In 1941, Kiyosi It\^o developed the formalism of stochastic calculus for inhomogeneous stochastic processes \cite{ito1944109, ito1951stochastic}, which generalises the theory of Brownian motion as given by Einstein, Smoluchowski, and Langevin to inhomogeneous diffusion \cite{bhattacharyay2025brownian}. It\^o's formal method of stochastic calculus, well-known as It\^o calculus, results in the same non-Fickian diffusion current that was shown by Chapman about twelve years in advance.  Diffusivity is naturally space-dependent in the presence of a temperature gradient, as it is proportional to the local temperature even when viscosity is temperature-independent. Moreover, the temperature dependence of diffusivity can be much more complex when viscosity itself is temperature-dependent. The diffusivity depends on particle concentration due to hydrodynamic effects, which must also be accounted for in the presence of concentration inhomogeneities, such as those arising from thermophoresis. The spatial dependence of diffusivity leads to additional transport, which can drive further concentration inhomogeneities beyond those driven by the solvation process. Explicit consideration of this stochastic process is, therefore, extremely important for the understanding of thermophoretic phenomena.

The Fick's law of diffusion in a homogeneous medium gives the Fickian diffusion current density (e.g., in one dimension along the $x$-direction), written as $j_F = - D (dC(x)/dx)$ where the diffusion constant is $D$. Diffusivity is a constant that indicates the space in which diffusion takes place is uniform. The non-Fickian diffusion current density $j_{NF}$ will arise in an inhomogeneous medium where the diffusivity $D(x)$ is a function of space. This current will arise as an additive part of the total diffusion current density $j = -dD(x)C(x)/dx$, where $j_{NF}=-C(x)(dD(x)/dx)$. At the origin of both the parts of the diffusion current $j_F$ and $j_{NF}$, the assumption of the local homogeneity and isotropy of space, which is a fundamental requirement of Einstein's theory of Brownian motion, remains intact \cite{bhattacharyay2025brownian}. This is why these are diffusion currents, unlike drift currents that break the local isotropy of space. From various perspectives, many groups have studied enhanced diffusion in active and other systems like enzymes and also in equilibrium Brownian systems, where this Chapman-It\^o non-Fickian diffusion current has been taken into account~\cite{eberhard2025force, jee2018catalytic, jee2018enzyme,weistuch2017spatiotemporal, agudo2018phoresis, bhattacharyay2025brownian, sharma2020conversion,sharma2023spontaneous, sharma2025entropic, sharma2026transport, bhattacharyay2019equilibrium, maniar2021random, bressloff2017temporal,bressloff2019protein,bhattacharyay2020generalization}. Works by Di Pu et al. have considered the Chapman-It\^o non-Fickian contribution in the field of thermophoresis ~\cite{pu2023colloid, pu2024mode, pu2024colloid, maier2020thermophoresis}. Seuwin et al. \cite{seuwin2024theory} have tried to explain the sign change in protein thermophoresis from a hydrogen-bonding perspective, to cite a few. Recent experimental studies have confirmed the presence of Non-Fickian diffusion within assemblies of proteins \cite{minarro2026non}. 

It is important to note that considering non-Fickian current in the context of thermophoresis is crucial because of the following reasons. The experimentally determined Soret coefficient $S_T(T)$ fixes the steady state concentration $C(T)$ as is schematically shown in Fig. 2(a). In all the processes where the $S_T(T)$ changes sign at an intermediate temperature, $C(T)$ has an extremum, and $dC(T)/dT$ is zero at that critical temperature $T^{*}$. This implies that the Fickian diffusion current will be zero at the temperature at which the concentration peak/trough occurs, as schematically shown in Fig. 2(b). In the non-equilibrium steady-state concentration, where the total current is zero, the solvation current must also pass through a zero at this temperature, where the Fickian current is zero, as schematically shown in Fig. 2(b). Moreover, the solvation current profile remains a mirror image of the Fickian current along the zero-current line in the non-equilibrium steady state. However, keep in mind that the solvation current is, in general, considered to be the cause of the whole effect, where the Fickian current matches it in magnitude everywhere in the sample in this steady state. Therefore, the zero crossing of the solvation current has nothing to do with the zero-crossing of the Fickian current, which results from the development of the steady state concentration profile as a response to the existence of the solvation current.

This implies a reversal in the direction of the solvation current relative to the same temperature gradient, even at the same concentrations on either side of the peak. To understand thermophoresis, therefore, one needs to answer how this solvation current reversal occurs in general. The current reversal can be explained quite easily by considering that the so-called solvation current consists of two components: the non-Fickian diffusion current and the other part, which would be the actual solvation contribution, where the latter has nothing to do with diffusive transport and is a drift term. In the steady state, the sum of the solvation and non-Fickian currents equals the Fickian diffusion current, which is the plausible picture that is schematically shown in Fig. 2(c). Moreover, one does not need to invent the non-Fickian diffusion current for this purpose, as it was established in the context of inhomogeneous diffusion on the basis of Einstein's theory of diffusion about a century ago. Moreover, in the standard theory of thermodiffusion where the diffusion current density is generally taken to be of the form $j^{diff}=-nD(\nabla c+\frac{k_T}{T}\nabla T)$ (as defined in the textbook - Nonequilibrium Statistical Physics: Linear Irreversible Processes, by No\"elle Pottier \cite{pottier2009nonequilibrium}), one can easily spot the non-Fickian diffusion current part which is proportional to the temperature gradient $\nabla T$. In the expression of the diffusion current density, $n$ is the total number density of molecules, $D$ is the diffusion coefficient, $c=n_A/(n_A+n_B)=n_A/n$ is the concentration of the species $A$ (say), and $Dk_T=D_T$ is the thermodiffusion coefficient. Therefore, this current stays hidden in the solvation current in thermophoresis when the entire attention is paid to the solvation mechanisms. The presence of Chapman's non-Fickian diffusion current in particle thermodynamics will inform the future understanding of the interfacial effects that give rise to a variety of solvation forces.   

Keeping this general picture in mind, the process of thermophoresis can be looked at as a phenomenon arising from the competition of three current densities. (1) The Fickian diffusion current, (2) the Non-Fickian diffusion current, and (3) the drift current due to solvation forces. In the present paper, we demonstrate the interplay of these three competing ingredients, resulting in Soret coefficients observed in three distinct experiments on aqueous solutions of Lysozyme, BLGA, and Poly-L-Lysine. We chose these three experimental results to avoid cases involving strong electrostatic forces at present; we will address them in the future. We show that non-Fickian diffusion is a major contributor to thermophoresis. For the class of systems under consideration, we show that the local density of water (assumed to have a low salt concentration), its thermal expansion coefficient, and a generic local pressure gradient can account for the observed $S_T(T)$ across all test cases. We also show that our model closely matches the experimentally observed equilibrium saturation concentration curve for Lysozyme, underscoring the important role of the non-Fickian current in both equilibrium and nonequilibrium phenomena. Moreover, we show that the theoretical model, built with these specific processes of thermophoresis in colloids in mind and involving a thermophilic-thermophobic transition, can also account for the general features of the Soret coefficient in other thermophoresis regimes where such a transition is absent. The model framework developed in this paper, based on results from kinetic theory and empirical inputs, captures fundamental aspects of phoretic transport driven by thermal gradients in a fluid and, once established, will greatly aid understanding of the interfacial effects that give rise to a variety of solvation forces in the future.

\begin{figure*}[t]
\centering
\begin{minipage}{0.31\textwidth}
    \centering
    \includegraphics[width=\textwidth]{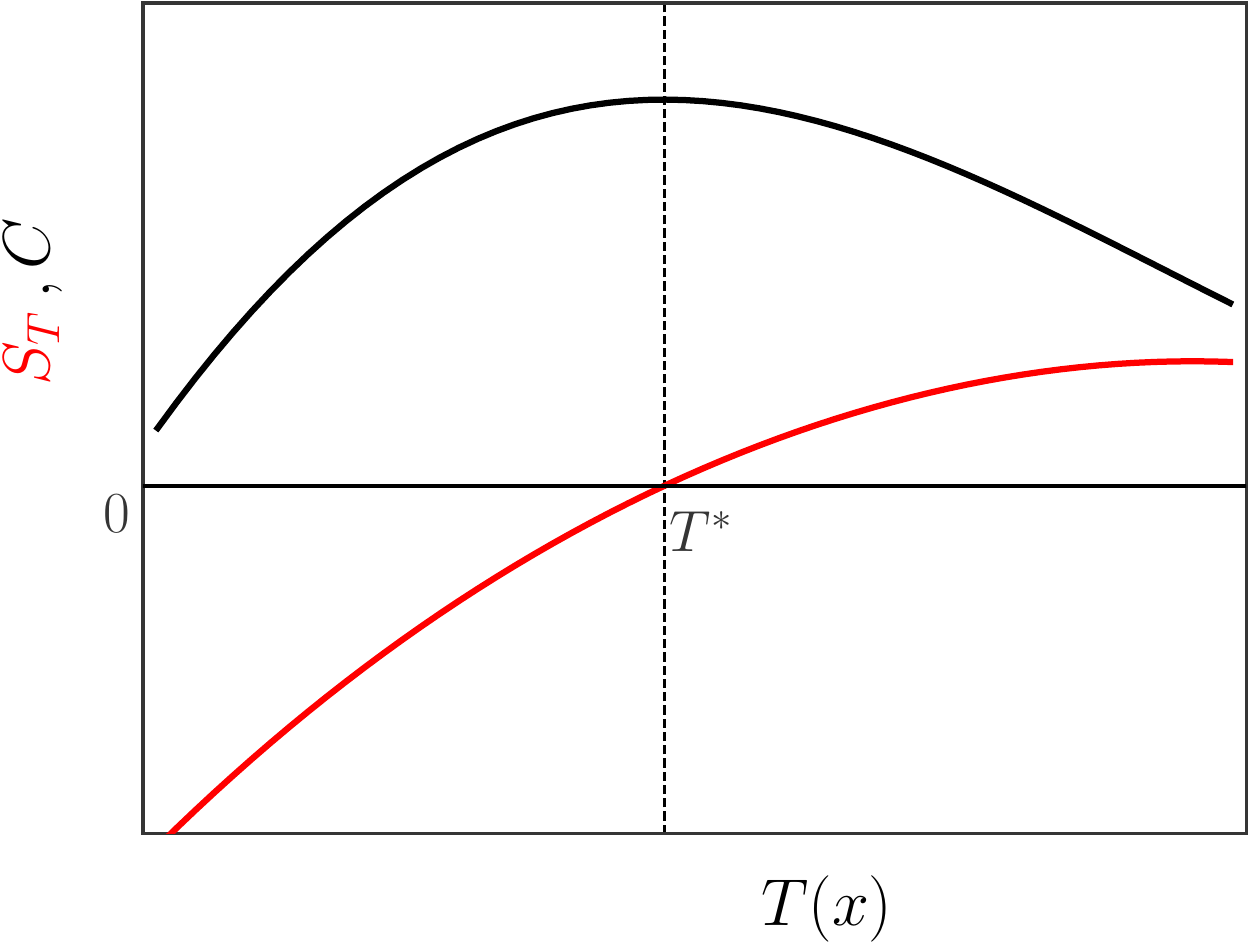}\\[0.25cm]
    {\large\textbf{(a)}}
\end{minipage}
\hspace{0.01\textwidth}
\begin{minipage}{0.31\textwidth}
    \centering
    \includegraphics[width=\textwidth]{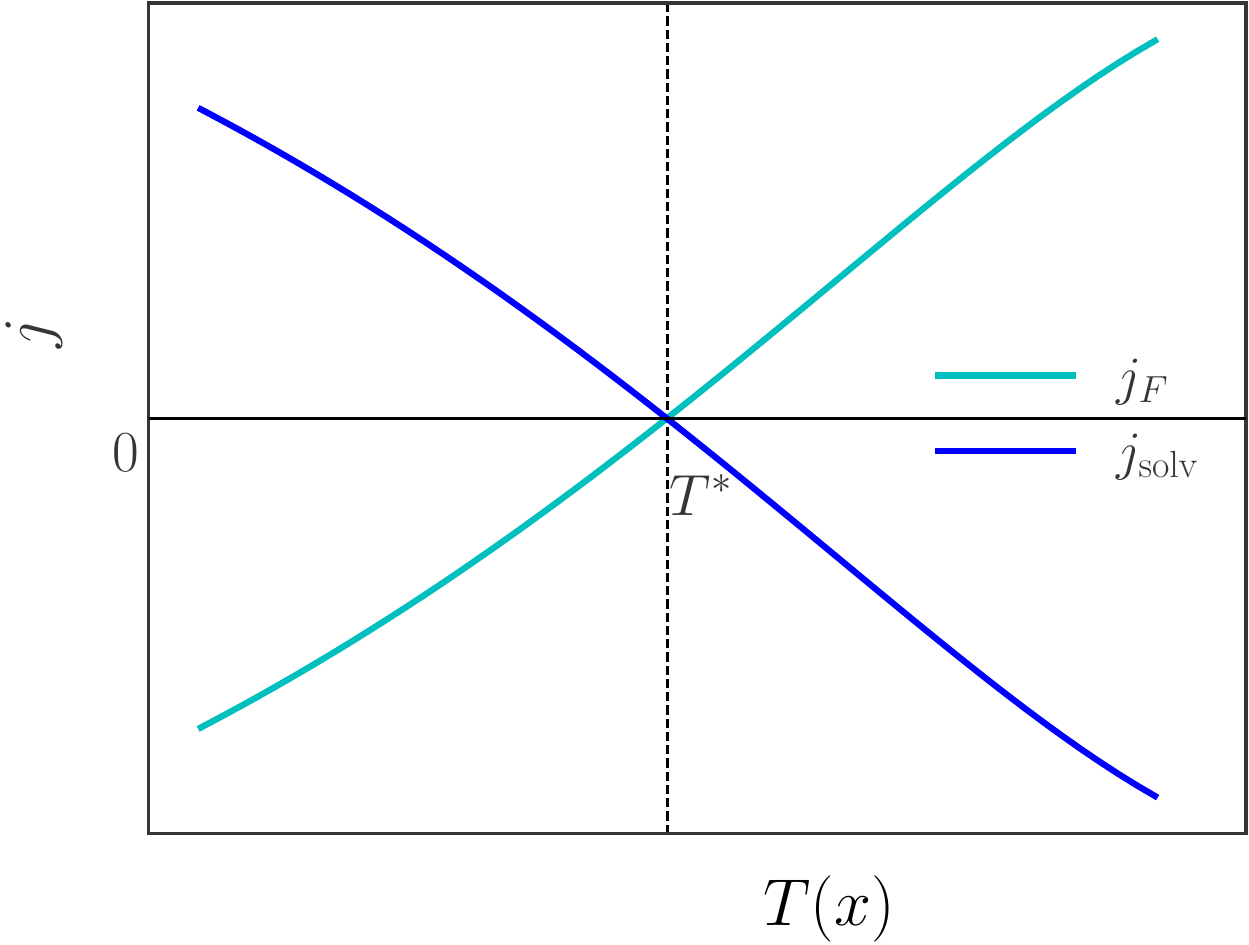}\\[0.25cm]
    {\large\textbf{(b)}}
\end{minipage}
\hspace{0.01\textwidth}
\begin{minipage}{0.31\textwidth}
    \centering
    \includegraphics[width=\textwidth]{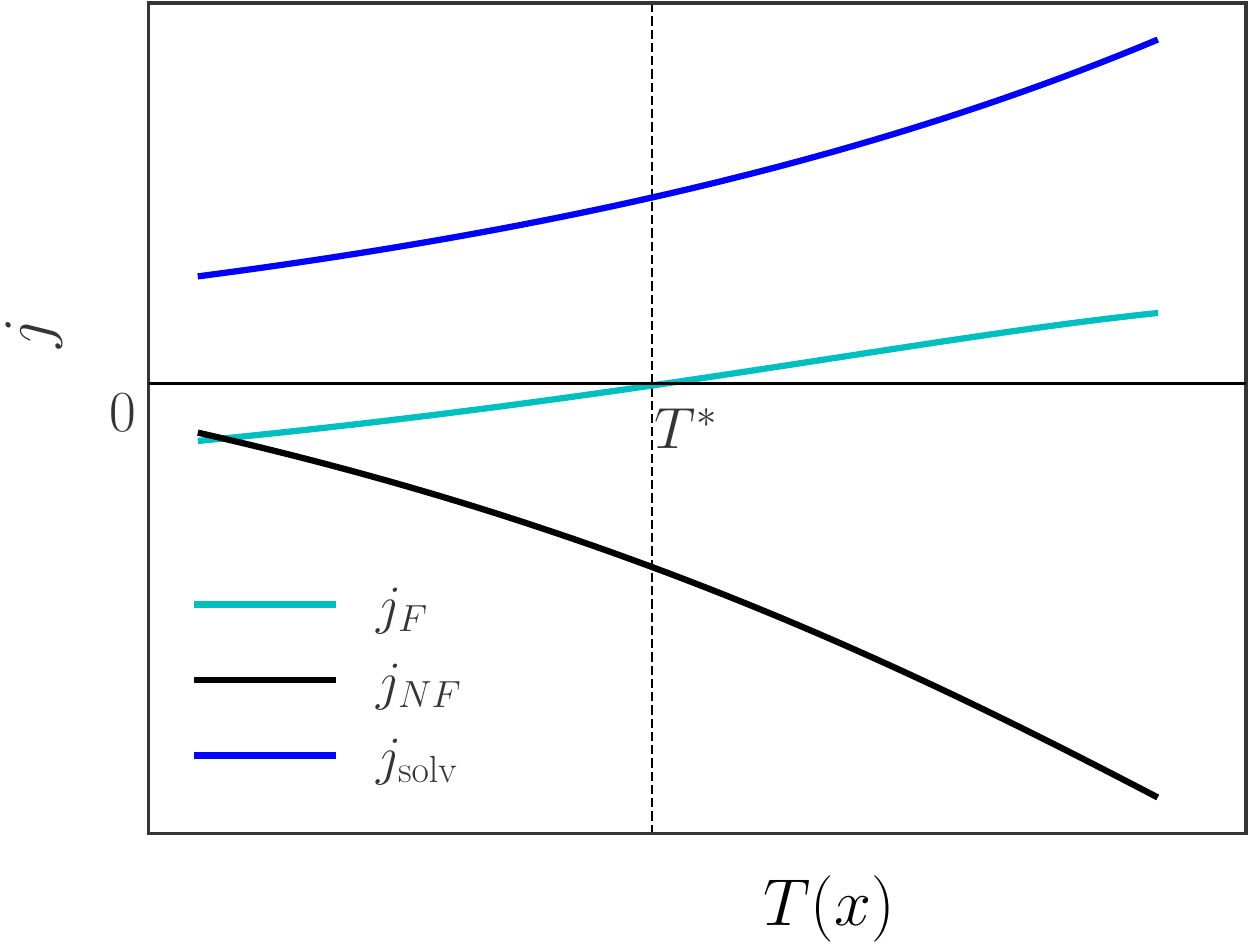}\\[0.25cm]
    {\large\textbf{(c)}}
\end{minipage}

\caption{
    Schematic illustration of (a) the Soret coefficient and corresponding concentration profile,
    (b) Two current competition scenerio: Fickian and solvation currents , and (c) three current competition scenerio:  Fickian,
    non-Fickian, and solvation currents. 
}
\label{fig:three_schematics}
\end{figure*}

\section*{ The model }
Before we write down the general expression for the current density that needs to be taken into account for the non-equilibrium process of thermophoresis, we need to understand the factors that can make the diffusivity of a colloidal particle in a solvent (water in the present paper with weak salt concentration)  space-dependent. Since we are working within the scope of Einstein's theory of diffusion, the local diffusivity of a colloidal particle follows the Stokes-Einstein relation 
\begin{equation}\label{eq:1}
D(x) = \frac{k_BT(x)}{6\pi\,\eta[T(x),C(x)]\,r}=\frac{k_B T(x)}{\Gamma[T(x),C(x)]}=D[T,C(T)],    
\end{equation}
where $k_B$ is the Boltzmann constant, $T(x)$ is the local temperature in a one-dimensional space in the presence of a temperature gradient $T=\tilde{g}x$, where $\tilde{g}$ is a constant. The local dynamic viscosity $\eta(T,C)$ is a function of space because it can depend on the local temperature as well as the particle concentration $C$ in general. Because interfacial proximity can alter damping through hydrodynamic effects, the dynamic viscosity may depend on particle concentration. In the Stokes-Einstein relation mentioned above, $r$ is the particle radius, which will be considered a constant. In the above relation, we specify the diffusivity as a general function of temperature and damping, $\Gamma(T, C)$. Finally, the diffusivity is, in general, an explicit function of temperature, since the particle concentration is.

One can now consider the general expression of the particle current density $j(T)$ in the context of thermophoresis (thermodiffusion) as
\begin{equation}
j(T) =-\frac{\partial}{\partial x}[C(T)D(T,C(T))]+\frac{D(T,C)}{k_{B}T}\, F_{sol}(T)\, C(T),
\label{eq:2}
\end{equation}
where $F_{sol}(T)$ is the temperature-dependent solvation force. The first term on the right-hand side of the above equation \eqref {eq:2} includes both the diffusion currents, Fickian and non-Fickian, as was given by Chapman. In the presence of a temperature gradient, considering that to be at the origin of the existence of the solvation force, a general expression for the solvation force is taken to be
\begin{equation}
F_{sol}(T) = a\,k_B\, f(T)\,\frac{dT}{dx},
\end{equation}
where $a$ is a constant, which depends on the other details of the physics behind the solvation force. In the above expression, $f(T)$ is a temperature-dependent function to be determined.

With this structure of the solvation force, the current density can finally be written as a function of the temperature gradient 
\begin{equation}
\begin{aligned}
j(T)= -\frac{dT}{dx}
\left[
\frac{\partial}{\partial T}\!\left(C(T)\,D(T,C)\right)
- a\,\frac{D(T,C)}{T}\, f(T)\, C(T)
\right].
\end{aligned}
\label{eq:4}
\end{equation}
The eq.{\eqref{eq:4}} results in a steady state non-equilibrium concentration profile of solutes corresponding to a zero current in the implicit form
\begin{equation}
\begin{aligned}
C(T)= \frac{\mathcal{N}}{D(T,C)} \exp\Big({a\int_{T_{0}}^{T}}\frac{f(T^{\prime})}{T^{\prime}}dT^{\prime}\Big),
\end{aligned}
\label{eq:5}
\end{equation}
where $\mathcal{N}$ is the normalization constant. With the stationary solute density corresponding to the zero current being known, the eq.{\eqref{eq:4}} can now be rewritten as
\begin{equation}\label{eq:6}
\begin{aligned}
j(T)
&= -DC\,\frac{dT}{dx}
\Bigg[
\left(
1 + \frac{C}{D}\left.\frac{\partial D}{\partial C}\right|_{T}
\right)\frac{1}{C}\frac{dC}{dT}
+ \frac{1}{D}\left.\frac{\partial D}{\partial T}\right|_{C}
- \frac{a}{T}\, f(T)
\Bigg].
\end{aligned}
\end{equation}
In the steady state, using the definition of the Soret coefficient $S_T\equiv -\frac{1}{C}\frac{dC}{dT}$, one now gets
\begin{equation}
\begin{aligned}
S_T
=
\frac{
\displaystyle -\frac{a}{T}\, f(T)
+ \displaystyle \frac{1}{D}\left.\frac{\partial D}{\partial T}\right|_{C}
}{
\displaystyle 1
+ \frac{C}{D}\left.\frac{\partial D}{\partial C}\right|_{T}
}.
\end{aligned}
\label{eq:7}
\end{equation}

\subsection{The general structure of the diffusivity}
The general form of the space-dependent diffusivity in the presence of a temperature and concentration gradient can be taken in a simple form as 
\begin{equation*}
D(T,C)=D_{0}(T)\,\Lambda(C),
\qquad
\Lambda(C)=\frac{1}{1+bC^{\beta}}.
\end{equation*}
The $D_{0}(T)$ represents the bare single-particle diffusivity, capturing the explicit temperature dependence arising from solute–solvent interactions (e.g. viscosity, hydration, and thermal agitation) in the dilute limit. The concentration-dependent factor of diffusivity $\Lambda(C)$ accounts for the dependence of diffusivity on crowding and ensuing hydrodynamic effects in general \cite{kuntz2004anomalous,ghosh2015anomalous,faucheux1994confined, kuntz2003anomalous}. The parameter $b$ sets the strength of the concentration dependence of the diffusivity. One may imagine that there exists a scale of concentration $C_0$, where $b=C_0^{-\beta}$. The dependence of the diffusivity on the local concentration is dominant for $C>C_0$.

The power-law structure used to define the concentration dependence of the diffusivity is a simplifying assumption to avoid invoking more than one parameter. The damping on a particle in water, in general, increases near an interface with a no-slip boundary condition as a function of the distance of the particle from the interface. The average radius $r$ of the free space around a particle at a concentration $C$ is proportional to $C^{-1/3}$. Based on the $r$-dependence of the hydrodynamic effects influencing diffusivity near an interface, the exponent $\beta$ should be determined from microscopic details. At the phenomenological level, in the present context, we expect the $\beta$ in this analysis not to vary much across similar systems considered as test cases.

The steady-state concentration can now be written using eq.\eqref{eq:5} as $C(T)\,\Lambda(C)=g[C(T)]$,
where
\begin{equation*}
g[C(T)]=\frac{1}{D_{0}(T)}
\exp\!\left[
a\int_{T_{0}}^{T}\frac{f(T')}{T'}\,dT'
\right].
\end{equation*}
Substituting the explicit form of $\Lambda(C)$, one gets $g[C(T)]=C/(1+bC^\beta)$.
The Soret coefficient for our model takes the form:
\begin{equation*}
S_T=\frac{\displaystyle-\frac{a}{T}\, f(T)+\frac{d}{dT}\ln D_0(T)}
{\displaystyle 1+\frac{d\ln \Lambda(C)}{d\ln C}}.
\end{equation*}
With the use of ${d\ln\Lambda(C)}/{d \ln C} = -{b\beta C^\beta}/{(1+bC^\beta)}$, the Soret coefficient can finally be written as a function of temperature and particle concentration as

\begin{equation}
S_T = \frac{\displaystyle -\frac{a}{T}\, f(T) +\frac{d}{dT}\ln D_0(T)}{\displaystyle 1-\frac{b\beta C^\beta}{1+bC^\beta}}.
\label{eq:8}
\end{equation}
 
At this stage, we can define the dominant asymptotic regimes of the concentration dependence of the diffusivity as (1) the weak concentration $bC^\beta << 1$, which should be the limit of thermodiffusion, and (2) the strong concentration $bC^\beta >> 1$, which is the limit of thermophoresis, where the corresponding Soret coefficients at the leading order are, respectively,
\begin{equation}
S_T\simeq -\frac{a}{T}\, f(T) +\frac{d}{dT}\ln D_0(T)
,
\qquad
bC^\beta\ll1,
\label{eq:9}
\end{equation}
and
\begin{equation}
S_T\simeq\frac{\displaystyle -\frac{a}{T}\, f(T) +
\frac{d}{dT}\ln D_0(T)}{ \displaystyle 1-\beta},
\qquad
bC^\beta\gg1.
\label{eq:10}
\end{equation}
These expressions of the Soret coefficients as explicit functions of temperature immediately indicate their dependence on the exponent $\beta$ in the latter case. The exponent $\beta$ being greater or less than one can alter the sign of the $S_T$. Note also from the above-mentioned expressions that if the solvation force parameter $a$ is independent of the particle size, so is the $S_T$. Moreover, the parameter $b$ does not explicitly show in the expression for $S_T$, as it corresponds to a concentration scale that is already accounted for in the qualitative consideration of the weak and strong regimes.

\subsection{Temperature dependence of the dynamic viscosity}
For a spherical particle of radius $r$, suspended in a solvent of dynamic viscosity $\eta(T)$, the temperature-dependent factor of diffusivity $D_{0}(T)$ is given by the
Stokes–Einstein relation,
\begin{equation}\label{eq:11}
D_{0}(T)=\frac{k_{B}T}{\Gamma_0(T)}
=\frac{k_{B}T}{6\pi\eta(T)r},
\end{equation}
Thermal gradients can alter the local solvent density, thereby modifying the local dynamic viscosity through a range of processes, e.g., from local hydrodynamics to hydrogen-bonding chemistry. The strength of this viscosity modification due to temperature gradient can be quantified by the solvent thermal expansion coefficient $\alpha_{w}(T)$, defined in terms of the temperature-dependent solvent density $\rho_w(T)$ as
\begin{equation}\label{eq:12}
\alpha_{w}(T)
=
-\frac{1}{\rho_{w}(T)}
\left.\frac{\partial \rho_{w}}{\partial T}\right|_{p},
\end{equation}
where the derivative is evaluated at constant pressure. Phenomenologically, we  take inspiration from \cite{iacopini2006macromolecular} and assume that the effective viscosity, as seen by the colloidal particle, can be written as
\begin{equation}\label{eq:13}
\eta(T)=\eta_{w}(T)-c\,\alpha_{w}(T),
\end{equation}
where $c$ is a system-specific free parameter for our model that controls the effect of the thermal expansion of the solvent on the local dynamic viscosity. At the origin of the correction term, there could be interactions between the particle's surface with the surrounding water, as well as the local hydrodynamics and maybe hydrogen bonding, which crucially depends on the orientation of water molecules \cite{nehme5448133microscale}. Normally, in the context of thermophoresis, such surface interactions are considered to generate a solvation force that drives particle concentration instabilities via a drift current. The diffusion is considered the process responsible for relaxation. However, the inclusion of the non-Fickian diffusion current opens the possibility of solvation mechanisms coupling to the diffusion process via a diffusivity gradient. Note that $\eta_{w}(T)$ stands for the temperature dependence of the viscosity of water obtained in equilibrium experiments, for example, in \cite{korson1969viscosity}.

\begin{figure}[t]
\centering
\begin{tikzpicture}
\node[anchor=south west, inner sep=0] (img) at (0,0)
    {\includegraphics[width=\textwidth]{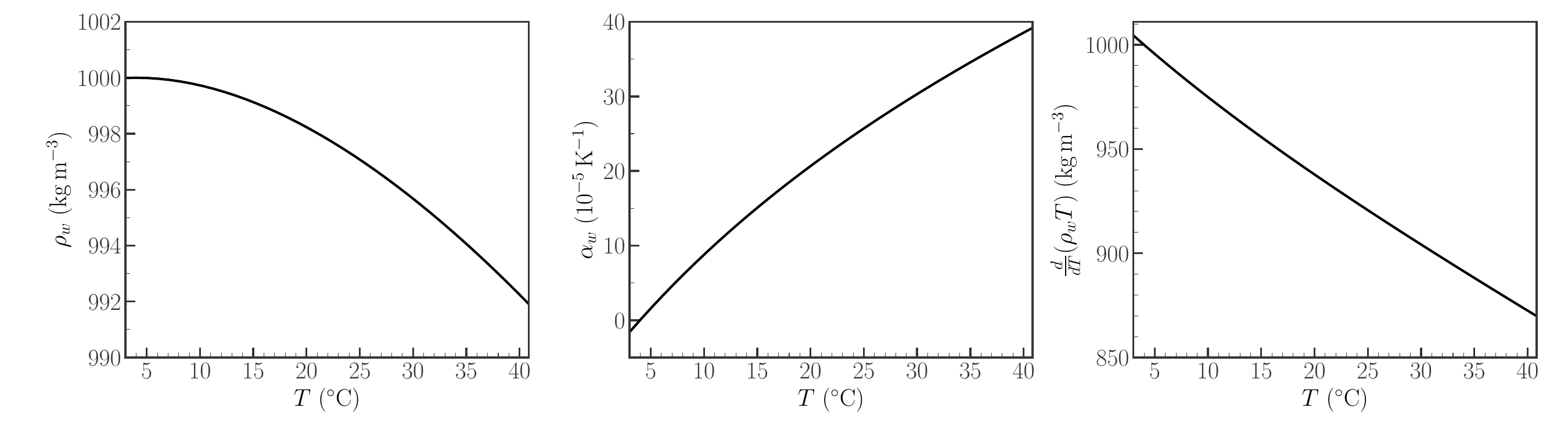}};
\begin{scope}[x={(img.south east)}, y={(img.north west)}]
    \node at (0.17,-0.12) {\large\textbf{(a)}};
    \node at (0.53,-0.12) {\large\textbf{(b)}};
    \node at (0.87,-0.12) {\large\textbf{(c)}};
\end{scope}
\end{tikzpicture}
\caption{Water properties as a function of temperature. Variation of (a) density $\rho_{w}$ of water, (b) its coefficient of thermal expansion $\alpha_{w}$, and (c) $\frac{d}{dT}\rho_{w}T$ as a function of temperature is shown. }
\label{fig_water_prop}
\end{figure}

Fig.~\ref{fig_water_prop} shows the typical variation of the density and the corresponding thermal expansion coefficient of water, where we take the standard empirical temperature dependence of the density of water to be
 
 \begin{equation}
 \label{eq:14}
 \rho_w(T) =\rho_w^{\max}\left[1-\frac{\left(T-277.1363\right)^2\left(T+15.7914\right)}
 {508929.2\,\left(T-205.0204\right)}\right],
 \
 273.15~\mathrm{K} \le T \le 313.15~\mathrm{K}.
 \end{equation}

This empirical dependence of water density on temperature (in K) at a pressure of 1 atm is reported in \cite{takenaka1990measurement, tilton1922accurate, tanaka2001recommended}. Moreover, the empirical relation of viscosity with temperature for a range of $273 - 363$ K at a pressure of 1 atm is taken to be
\begin{equation}\label{eq:15}
\begin{aligned}
\eta_{w}(T)= A \exp\big({\frac{B}{T-\theta}}\big),
\end{aligned}
\end{equation}
from the reference \cite{likhachev2003dependence}.
In this empirical relation, $A=2.4152 \times10^{-5}$\,Pa-s, $B= 570.46$\,K and $\theta=139.86$\,K and $T$ is in K. In the present paper, we will use $\rho_w(T)$ and $\eta_w(T)$ as mentioned in the equation \eqref{eq:14} and \eqref{eq:15}, respectively.


\subsection{The solvation force}
Having developed the model for the diffusive part, let us examine the structure of the function $f(T)$ in the solvation force. The simplest way to model the solvation force causing the drift current is to make it proportional to the negative gradient of the local pressure on a particle, i.e. $F_{sol}(T)\propto -dP/dx = -(dP/dT)(dT/dx)$. Now take the reasonable equation of state that $P(T)= R\, \rho_{w}(T) \, T$, where $R$ is a constant. Fig.~\ref{fig_water_prop}(c) clearly indicates that, within the region of interest in temperature, i.e., 5 to 40 \degree C, the derivative $dP/dT$ is predominantly linear and negative. Therefore, one may assume the solvation force is generally of the structure $F_{sol}(T)\propto T(dT/dx)$ giving $f(T)\propto T$. In this paper, we set $f(T)=T$ as the constant $a$ is already present in the expression of the solvation force. The local equation of state for water in the vicinity of the particle is being considered proportional to the density of water and its local temperature, keeping in mind the nature of the temperature dependence of the density of water in the relevant range of temperatures.

Let us explain here in brief as to why we are intending to look into the origin of the solvation force $F_{sol}(T)\propto T(dT/dx)$, in the present context, as something primarily local pressure-driven than local shear stress-driven drag on a phoretic particle, which is generally assumed spherical in microscopic theoretical models. The primary reasons for these are as follows: (a) It is the local pressure gradient that must develop first due to broken symmetry, for example, the presence of a temperature gradient, in a previously hydrostatic steady state of a fluid, and only then do the shear stresses (off-diagonal elements of the pressure tensor) arise as the fluid starts flowing, generating some velocity gradient. (b) Even under Stokes flow at low Reynolds numbers, where the flowing fluid drags the particle, the components of the drag force on the fluid due to local pressure and shear stresses are comparable. (c) In the present phenomenological modelling approach, the local reduction of dynamic viscosity (Eq. (13), and parameter $c$) is a crucial ingredient that accounts for the curvature of the experimental Soret coefficient curves by considering non-Fickian currents. Under such an approach, the local flow of the fluid around the phoretic particle can deviate substantially from Stokes' flow pattern, and the pressure-driven force on the particle can be the dominant contributor compared with shear-stress forces.

Our perspective here differs from that advocated by Piazza and others in their works, as noted in the references \cite{piazza2008thermophoresis,piazza2004thermal,piazza2008thermophoresis2, iacopini2003thermophoresis, parola2005microscopic}. These works, primarily from the Piazza and coworkers, are motivated by the earlier work of Ruckenstein \cite{ruckenstein1981can}. Ruckenstein, in his proposal, has suggested that, despite being a weak influence on a solid-liquid interface, thermophoretic forces can have a partial origin in the direction of decreasing interfacial tension, as arises in the Marangoni effect at a liquid-liquid interface. It is important to note that, in his paper \cite{ruckenstein1981can}, Ruckenstein had particularly stressed the fact that such interfacial tension-driven forces on a particle (liquid-solid interface) are, in general, small in nature. Our approach to modelling in the present paper differs fundamentally from that of Piazza and coworkers, not only in treating the solvation force as a function of the local pressure gradient but also in considering the non-Fickian diffusion current alongside the solvation and Fickian currents. When a model is based on the effects of local thermal expansion of the fluid (water) on diffusion, the corresponding density gradients of the fluid cannot be neglected. This means, in the presence of a temperature gradient, the local pressure gradient cannot be neglected. Note also the fact that the temperature-dependent form of the solvation force, as taken by us $F_{sol}(T)\propto T(dT/dx)$, is actually the same as considered by Ruckenstein and others \cite{iacopini2003thermophoresis, ruckenstein1981can}. It is only the physical origin of such a form of the solvation force that we are stressing, the local pressure gradient generated, where Piazza et al. advocate that it originates from interfacial stresses.

\section*{Results: Comparison with experiment}
\begin{figure}[h]
\centering
\begin{tikzpicture}
\node[anchor=south west, inner sep=0] (img) at (0,0)
    {\includegraphics[width=\textwidth]{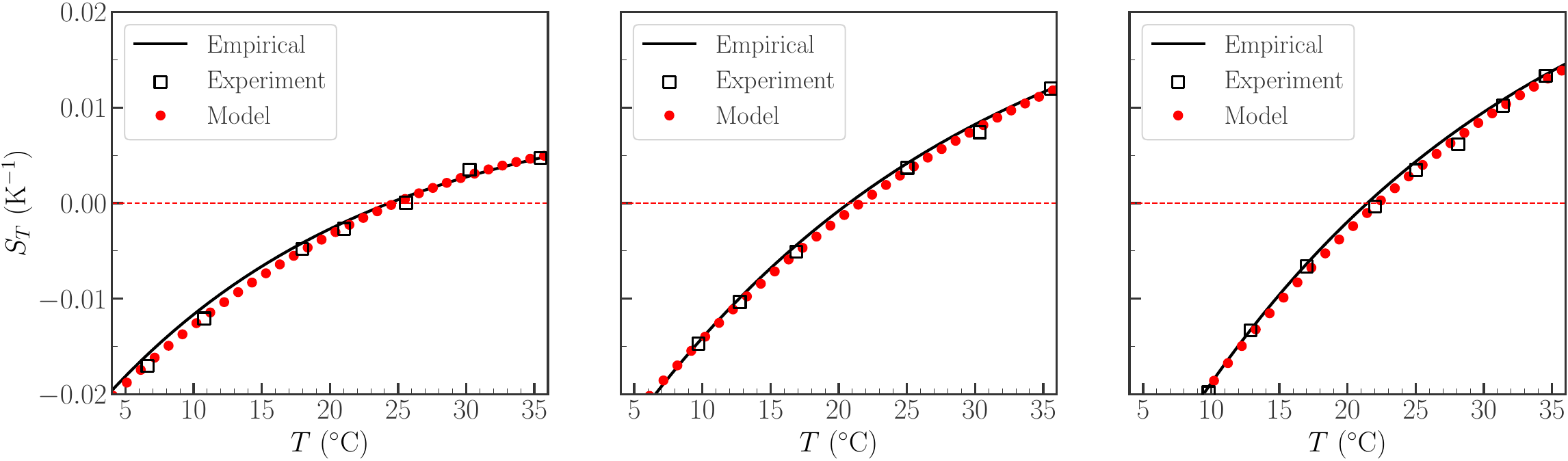}};
\begin{scope}[x={(img.south east)}, y={(img.north west)}]
    \node at (0.17,-0.12) {\large\textbf{(a) }\,\normalsize Lysozyme};
    \node at (0.53,-0.12) {\large\textbf{(b)}\,\normalsize BLGA};
    \node at (0.84,-0.12) {\large\textbf{(c)}\,\normalsize Poly-L-Lysine};
\end{scope}
\end{tikzpicture}
\caption{Soret coefficient comparison between experiment data, empirical form, and model predictions shown for Lysozyme, BLGA, and Poly-L-Lysine. The experimental data shown have been extracted from Fig.1 of the reference \cite{iacopini2006macromolecular} using an online data-digitisation tool (PlotDigitizer \cite{PlotDigitizer}).  }
\label{fig_protein_theo_exp_comp}
\end{figure}

Let us compare our model with the curves fitted to the experimental data for hen egg-white lysozyme and other polypeptide solutions reported in ref. \cite{iacopini2006macromolecular, iacopini2003thermophoresis, piazza2004thermophoresis}. The empirical fitting function chosen to fit the experimentally obtained data for the Soret coefficient is
\begin{equation}\label{eq:16}
S_T^\prime
=
S^{\infty}_T
\left[1-\exp\,\left(\frac{T^{*}-T}{T_{0}}\right)\right],
\end{equation}
 where $S^{\infty}_T$ is the high-temperature thermophobic limit, $T^{*}$ is the temperature at which $S^\prime_{T}$ switches sign, and $T_{0}$ denotes the scale of exponential growth. Note that the transition temperature $T^{*}$ is poorly sensitive to salt addition and Lysozyme charge \cite{iacopini2006macromolecular,iacopini2003thermophoresis, piazza2004thermophoresis}.  
The above-mentioned empirical function for the Soret coefficient in these systems is applicable over a wide range of pH and ionic strength, indicating that neither parameter significantly affects the ansatz. Our model at present does not include explicit contributions from pH or ionic strength; beyond this limit, they are accounted for through the parameters $c$ and $a$. Thus, our model is well-suited for comparison with experimental data that, in general, matches the empirical Soret coefficient of eq.\eqref{eq:16}. The value of the three free parameters of $S_T^\prime$ in eq.\eqref{eq:16} and the corresponding three free parameters in the expression of $S_T$ in our model, which make the curves match $S^\prime_T$ that matches the experimental data, are shown in Table \ref{tab:ST_compare}. Figure \ref{fig_protein_theo_exp_comp} compares curves with the $S_T$ from our model plotted as red dots. The experiments on Lysozyme, whose data are presented here, were carried out in a sodium acetate buffer at
pH$ = 4.65$, with a Lysozyme concentration $7~\mathrm{g\,L^{-1}}$ and $0.4~\mathrm{M}$ NaCl (corresponding to $2.3\%~\mathrm{w/v}$)
\cite{piazza2004thermophoresis, iacopini2003thermophoresis, iacopini2006macromolecular}.
The experiments on BLGA and Poly-L-Lysine were performed at pH$= 7.0$, with BLGA at a concentration $13~\mathrm{g\,L^{-1}}$ in the presence of $50~\mathrm{mM}$ NaCl, and $50~\mathrm{kDa}$ Poly-L-Lysine at concentration $5.4~\mathrm{g\,L^{-1}}$ with $100~\mathrm{mM}$ NaCl.

\renewcommand{\arraystretch}{1.3}
\setlength{\tabcolsep}{8pt}

\begin{table}[htbp]
\centering

\begin{minipage}[t]{0.48\textwidth}
\centering
\begin{tabular}{lccc}
\hline\hline
Sample & $S_T^{\infty}$ (K$^{-1}$) & $T^{*}$ (K) & $T_0$ (K) \\
\hline
Lysozyme & $0.0111 \pm 0.0015$ & $297.5 \pm 0.3$ & $20 \pm 2$ \\
BLGA & $0.0275 \pm 0.0035$ & $293.9 \pm 0.3$ & $26 \pm 3$ \\
Poly-L-lysine & $0.034 \pm 0.006$ & $294.6 \pm 0.3$ & $26 \pm 4$ \\
\hline\hline
\end{tabular}
\end{minipage}
\hfill
\begin{minipage}[t]{0.48\textwidth}
\centering
\begin{tabular}{lccc}
\hline\hline
Sample & $a$$(K^{-1})$ & $\beta$ & $c\,(\mathrm{Pa}\,\mathrm{s}\,\mathrm{K})$
 \\
\hline
Lysozyme & $0.0314$ & $1.265$ & $0.275$ \\
BLGA & $0.0303$ & $1.232$ & $0.183$ \\
Poly-L-Lysine & $0.0303$ & $1.179$ & $0.195$ \\
\hline\hline
\end{tabular}
\end{minipage}


\caption{Parameters for the temperature dependence of the Soret coefficient $S_T^\prime$ \cite{iacopini2006macromolecular} [left] and $S_T$ from our model [right] for different polypeptide solutions obtained via least-squares minimisation with respect to experimental data; the details can be found in the Appendix.}
\label{tab:ST_compare}

\end{table}

The Soret coefficient $S_T$ of our model in the strong concentration regime is
\begin{equation}\label{eq:17}
S_T = \frac{1}{1-\beta} \left(\frac{d}{dT}\log D_{0}(T)\right)-\frac{a}{1-\beta},
\end{equation}

\begin{figure}[t]
\centering
\begin{tikzpicture}
\node[anchor=south west, inner sep=0] (img) at (0,0)
    {\includegraphics[width=\textwidth]{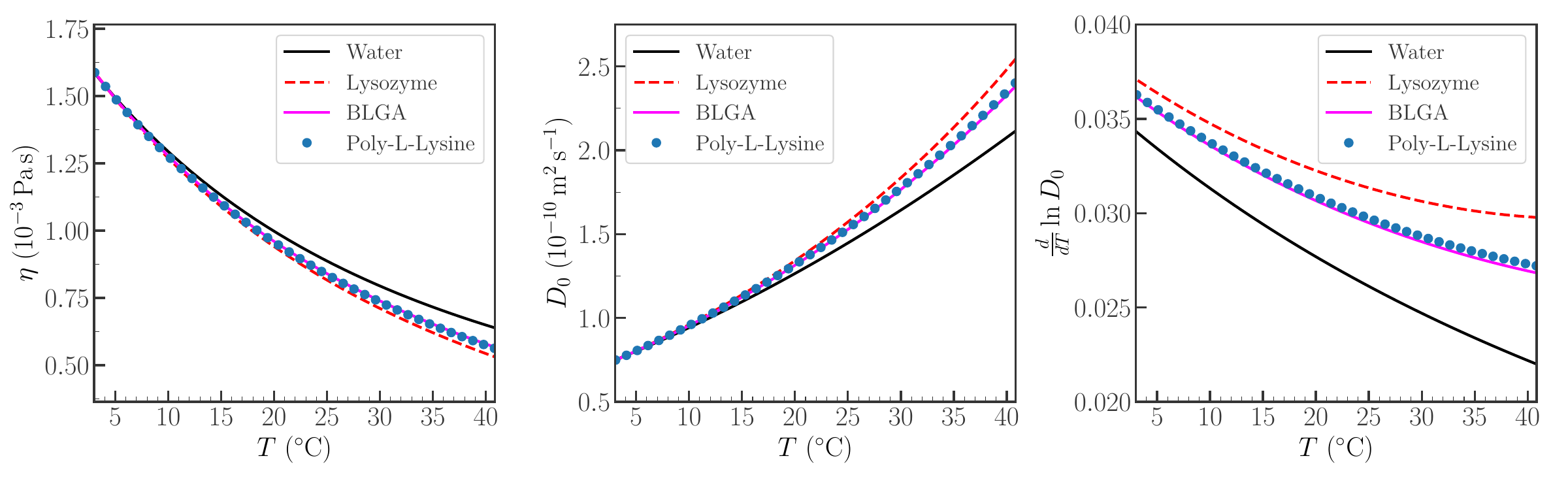}};
\begin{scope}[x={(img.south east)}, y={(img.north west)}]
    \node at (0.17,-0.12) {\large\textbf{(a)}};
    \node at (0.53,-0.12) {\large\textbf{(b)}};
    \node at (0.87,-0.12) {\large\textbf{(c)}};
\end{scope}
\end{tikzpicture}
\caption{The figure (a) shows the variation of viscosity and (b) that of $D_0(T)$ with temperature, where (c) shows the variation of the temperature gradient of the logarithm of $D_0(T)$ with temperature. The protein particle radius taken is 1.7 nm \cite{ iacopini2003thermophoresis, piazza2004thermophoresis}. }
\label{fig_visc_diff_logdiff}
\end{figure}

which clearly indicates that the part of the solvation force that contributes to the drift current produces only a vertical shift in the curve and does not contribute to the curvature. However, we realised that the parameter $c$ in the expression for dynamic viscosity plays a dominant role in determining the correct curvature. One may recall, as we have already alluded to, that this parameter $c$ may embody the coupling between the solvation mechanisms and the diffusion. This physics is absent when the non-Fickian diffusion term is omitted from the model. Note that the exponent $\beta$ sets a scale for the $S_T$ as well, besides controlling the sign according to our model.

To elucidate the major role of the parameter $c$ in determining the shape of the $S_T$ curve, we show in Fig.~\ref{fig_visc_diff_logdiff}(a) and ~\ref{fig_visc_diff_logdiff}(b) the dependence of $\eta(T)$ and the corresponding $D_0(T)$ on temperature for all three cases. We have also plotted $\eta_w(T)$ (water) in these figures for a visual comparison. In Fig.~\ref{fig_visc_diff_logdiff}(c), we plot the derivative of $\ln D_0(T)$ with respect to temperature for all cases, including water. Fig.~\ref{fig_visc_diff_logdiff}(c) makes it clear that the first term on the right-hand side of eq.\eqref {eq:17} is negative, while the second term is positive because $\beta>1$. 
From these curves, it is clear that while BLGA and Poly-L-Lysine are similar systems, Lysozyme is somewhat a different system in nature. Moreover, while the phoretic activity of Lysozyme is substantially dependent on the thermal expansion of water ($c=0.275$), the BLGA and Poly-L-Lysine show a relatively reduced contribution of the correction to dynamic viscosity depending on the local thermal expansion coefficient of water $\alpha_w(T)$ with a $c\simeq 0.183$ and $c\simeq 0.195$, respectively. Since $\alpha_w$ is practically positive in the considered temperature range, the coupling between the solvation mechanism and diffusion in this temperature range, due to a positive $c$, effectively reduces viscosity in the solvation layer. The physics of this, which the present model indicates, hides in the actual solvation mechanism at play, which needs to be understood.

\begin{figure}[h]
\centering
\begin{tikzpicture}
\node[anchor=south west, inner sep=0] (img) at (0,0)
    {\includegraphics[width=\textwidth]{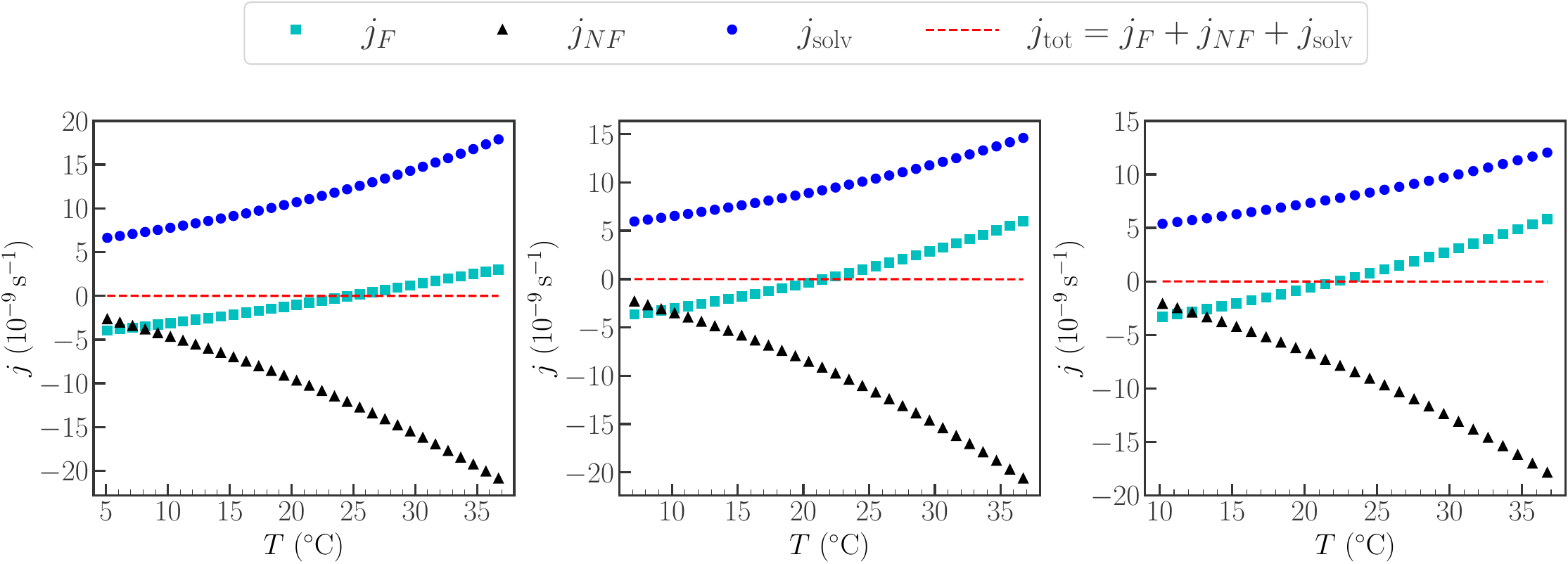}}; 
\begin{scope}[x={(img.south east)}, y={(img.north west)}]
    \node at (0.17,-0.12) {\large\textbf{(a)}\, Lysozyme};
    \node at (0.53,-0.12) {\large\textbf{(b)}\, BLGA};
    \node at (0.87,-0.12) {\large\textbf{(c)}\,Poly-L-Lysine};
\end{scope}
\end{tikzpicture}
\caption{Competition between Fickian, non-Fickian diffusion currents and solvation force-driven drift shown for Lysozyme, BLGA, and Poly-L-Lysine.}
\label{fig_current_comp}
\end{figure}

\begin{figure}[h]
\centering
\begin{tikzpicture}
\node[anchor=south west, inner sep=0] (img) at (0,0)
    {\includegraphics[width=\textwidth]{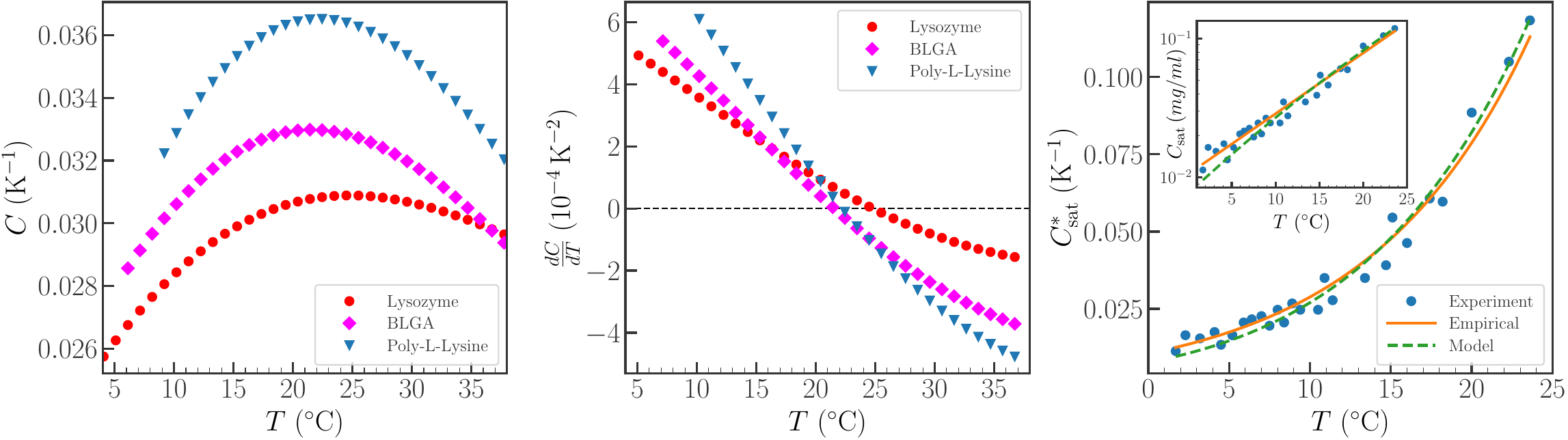}};
\begin{scope}[x={(img.south east)}, y={(img.north west)}]
    \node at (0.17,-0.12) {\large\textbf{(a)}\, };
    \node at (0.53,-0.12) {\large\textbf{(b)}\, };
    \node at (0.87,-0.12) {\large\textbf{(c)}\,};
\end{scope}
\end{tikzpicture}
\caption{Figure shows variation of (a) concentration (normalised), (b) derivative vs temperature plot obtained from our model for thermophoresis of Lysozyme, BLGA, Poly-L-Lysine, respectively. In (c), we display the Comparison of normalised saturation concentration ($C^*_{sat}$) with temperature obtained from equilibrium experiment and the model, shown for chicken egg-white Lysozyme in the main body figure~\cite{forsythe1999tetragonal,cacioppo1991solubility,iacopini2003thermophoresis, carpineti2004metastability, piazza2004thermophoresis}. The upper-right inset (a semi-log plot) shows the variation in saturated concentration (un-normalised) with temperature.}
\label{fig_pred}
\end{figure}

To obtain a clearer estimate of the competing Fickian, non-Fickian diffusion currents and the solvation force-driven drift, which are in balance in the steady state, we show plots of these currents for each case under consideration against temperature in Fig.~\ref{fig_current_comp}. In these figures, one can see that the BLGA and Poly-L-lysine have very similar current density profiles, whereas Lysozyme shows a distinct profile, with the rise in solvation current with temperature occurring more at the higher-temperature end. Moreover, the dominant role in the competition of currents is played between the non-Fickian current and the drift induced by the solvation forces. The Fickian diffusion current is smaller in magnitude than the other two currents; however, it reverses direction when the corresponding concentration gradient changes sign with temperature, thereby changing the spatial concentration gradient in the presence of a linear temperature gradient. This makes the Fickian current cross zero near about $22 \degree\text{C}$ for BLGA and Poly-L-Lysine, and near $25 \degree\text{C}$ for Lysozyme, as shown in Fig.~\ref{fig_pred}(a).

From ref.~\cite{iacopini2003thermophoresis, piazza2003thermal}, we find that a temperature difference of $\Delta T \sim 1\,\mathrm{K}$ is imposed over a plate length scale $\Delta x = 0.7\,\mathrm{mm}$, yielding a temperature gradient $\Delta T / \Delta x \approx 10^{3}\,\mathrm{K\,m^{-1}}$. We have taken this value for $dT/dx$ in our calculations of current shown in Fig.~\ref{fig_current_comp}.
In all these figures, the total current is zero, as shown by the dotted central line. Fig.~\ref{fig_pred}(a) and Fig.~\ref{fig_pred}(b) show the plot of concentration $C(T)$ and $dC(T)/dT$ against $T$ to present a visual guide as to where the Fickian current crosses the zero. Note that we have plotted concentrations only within the experiment's temperature range. The temperature at the distribution's maximum corresponds to the transition temperature, which must happen. The transition temperature is the one at which the Soret coefficient vanishes. This can also be confirmed by comparing Fig.~\ref{fig_pred}(b), where the variation of the temperature derivative of Concentration with temperature is shown.

The temperature-dependent particle concentration in our model in the strong concentration regime is
\begin{equation}\label{eq:eq}
\begin{aligned}
C(T)= \mathcal{N^{\prime}}\Bigg[\frac{1}{D_{0}(T)} \exp\Big({a\,T}\Big)\Bigg]^{1/1-\beta}.
\end{aligned}
\end{equation}

Concentration-dependent cluster formation in Lysozyme solutions and non-Fickian diffusion in proteins have been observed in experiments \cite{zhang2024effective, minarro2026non}. In refs.~\cite{forsythe1999tetragonal,iacopini2003thermophoresis,
pusey1988method,pusey1991micro}, Pusey and coworkers measured the saturation concentration $C_{\mathrm{sat}}$ of Lysozyme solutions in equilibrium with tetragonal Lysozyme crystals, which form a stable ordered phase below $25\,^{\circ}\mathrm{C}$. Measurements were performed for various pH values, bath temperatures, and salt concentrations. The part of the experiment relevant to our discussion is carried out at a fixed pH and temperature at a fixed (low) salt concentration. In this protocol, two columns containing Lysozyme crystals are maintained in contact with a thermal bath at a temperature $T$. A solution consisting of buffer, water, NaCl, and Lysozyme is passed through each column from one end. In the first column, an undersaturated solution is used, leading to dissolution of the solid Lysozyme until the concentration of the solution exiting the column reaches a steady saturation value $C_{\mathrm{sat}}$. In the second column, a supersaturated solution is introduced, causing Lysozyme from the solution to deposit onto the crystal until the same equilibrium saturation concentration $C_{\mathrm{sat}}$ is reached. In both cases, the system relaxes to a unique concentration at which dissolved Lysozyme and solid crystals coexist in equilibrium, and this value is recorded as $C_{\mathrm{sat}}$. 

The experiment is then repeated at a different bath temperature, while keeping the pH and buffer properties unchanged. In this way, the temperature dependence of the saturation concentration is obtained. Experimentally, this dependence is well described by the empirical form given by Piazza et al. as $C_{\mathrm{sat}}(T) = C_{0}\exp\!\left(\frac{T}{T_{1}}\right),$
with $T_{1}$ in the range $8-13\,{\mathrm{K}}$ ~\cite{forsythe1999tetragonal,cacioppo1991solubility,iacopini2003thermophoresis, carpineti2004metastability, piazza2004thermophoresis}. Importantly, no temperature gradient is present in these experiments, so it probes an equilibrium distribution rather than a thermophoretic steady state. The normalised experimental data of Pusey et al. at pH of 4.6, 3\% NaCl in solution, and that of empirical form are compared with the normalised equilibrium saturation distribution called $C^{*}_{sat}$ predicted by our model, as shown in Fig.~\ref{fig_pred}(c). We have chosen this data because it is closest to the thermophoresis experiment on Lysozyme, which was performed at pH 4.65 and about 2.33\% NaCl. On comparison, we find an excellent match between our model and both data sets. The value of \( T_{1} \) found for the empirical form is \( T_{1} = 9.95\,\mathrm{K} \). The inset in the same figure shows a semi-logarithmic plot of the saturation concentration \( C_{\mathrm{sat}} \) (in units of g/mL) from the experimental data, the empirical form, and the prediction of our model. Note that we have set $a=0$ and $c=0$ in the strong concentration regime ($\beta$ is the same as that of Lysozyme in Fig.~\ref{fig_pred}(a)) for our graph in Fig.~\ref{fig_pred}(c).

It is important to note that, for stationary equilibrium distributions at constant temperature in a saturated solution, there apparently is no dominant role for diffusive transport, since the concentration is stationary. However, the diffusive transport has played its role during the transition to the equilibrium distribution. The fact that the equilibrium concentration depends on $D_0(T)$ and $\beta$ as in equation \eqref{eq:eq} is a testament to the reality that the system's Fokker-Planck equation is nonlinear in concentration, since the diffusivity is concentration-dependent. This equilibrium distribution is the result of an It\^o-process, which is characterised by the presence of a non-Fickian diffusion current, as was identified by Chapman. In this sense, matching the experimental results for the equilibrium distributions with those of the It\^o-process is significant for understanding the indispensability of non-Fickian diffusion currents.

\begin{figure}[h]
\centering
\begin{tikzpicture}
\node[anchor=south west, inner sep=0] (img) at (0,0)
    {\includegraphics[width=\textwidth]{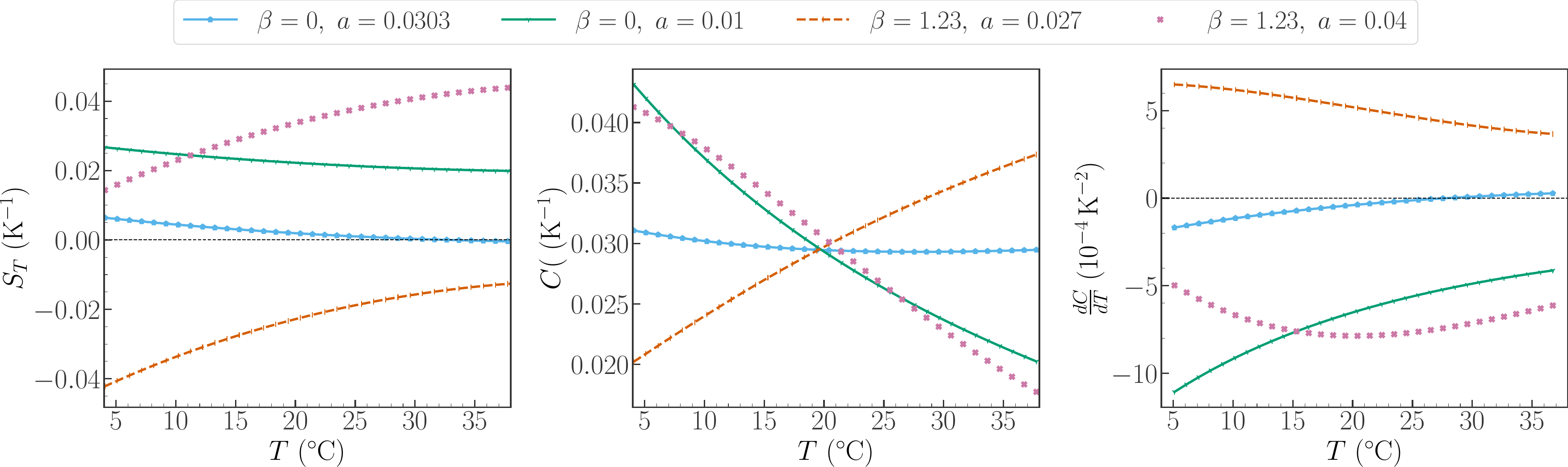}}; 
\begin{scope}[x={(img.south east)}, y={(img.north west)}]
    \node at (0.17,-0.12) {\large\textbf{(a)}\, };
    \node at (0.53,-0.12) {\large\textbf{(b)}\,};
    \node at (0.87,-0.12) {\large\textbf{(c)}\, };
\end{scope}
\end{tikzpicture}
\caption{Soret coefficient obtained from the model keeping $c=0.275$ fixed and varying $\beta$ and $a$ shown in $(a)$ and  variation of the corresponding concentration and its temperature derivative  with temperature is shown in $(b)$ and $(c)$  respectively for the temperature range of $4\,^{\circ}\mathrm{C}$
 to $38\,^{\circ}\mathrm{C}$ 
. }
\label{fig_pred_l}
\end{figure}

In Fig.~\ref{fig_pred_l}, we show some plots of the Soret coefficient and respective concentration obtained from our model for low concentration and high concentration diffusion regimes. Our model clearly shows that there can exist regimes in which the system remains completely thermophobic or thermophilic, or undergoes a transition from one regime to another, depending on the solvation parameter $a$, the exponent $\beta$, or the parameter $c$. Note that when the diffusivity has no explicit concentration-dependence ($\beta=0$), the Soret coefficient becomes $S_{T}= -a +\frac{d}{dT}\log D_{0}(T)$. The first term `$a$' is the solvation term whose strength will be determined by the interaction of the particle surface with the fluid.  When the contribution from the solvation term is negligible in comparison to the diffusive term, the Soret coefficient remains positive throughout in accordance with our model, as shown by parameter $a =0.01 $ in Fig.~\ref{fig_pred_l}(a). This kind of behaviour was quantitatively seen by Ludwig and Soret in their experiment in \cite{ludwig1856difusion, soret1879etat, platten} on thermodiffusion, but thermodiffusion needs a different model than the present one, even while taking into consideration the non-Fickian current. 

When the non-Fickian diffusive and solvation terms become comparable to the Fickian term at some temperature, there can be a transition from thermophobic to thermophilic nature as shown in Fig.~\ref{fig_pred_l}(a) using in the plot with $a=0.0303$. This transition is further confirmed by the concentration–temperature profile in Fig.~\ref{fig_pred_l}(b), where the transition temperature corresponds to a local minimum in the concentration. Consistently, in Fig.~\ref{fig_pred_l}(c), the derivative of the concentration with respect to temperature crosses zero at the same temperature. Moreover, it is worth pointing out that there also exists a regime (strong concentration-dependent diffusivity, here for $\beta =1.23$) in which the Soret coefficient can remain negative/positive throughout, as shown in Fig.~\ref{fig_pred_l}(a). One can shift these curves vertically by varying the solvation parameter $a$ to observe a transition. The concentration and its temperature derivative corresponding to the parameters are displayed in Fig.~\ref{fig_pred_l}(b) and Fig.~\ref{fig_pred_l}(c), respectively.

\section*{Discussion}
The model we consider in the present paper is based on Fickian and non-Fickian diffusion currents, as first proposed by Chapman, who generalised Einstein's theory of diffusion to inhomogeneous fields. The same currents appear in the formal theory of stochastic calculus as proposed by It\^o. In the present analysis, we have sought to incorporate a broader scope of non-Fickian current into our model, in which diffusivity can vary spatially due to temperature and particle concentration gradients. Both these gradients are extremely important in making the diffusivity depend on coordinates. The concentration dependence of the diffusivity arises from hydrodynamic effects that become substantial for a diffusing particle near an interface. In higher-concentration regions, the hydrodynamics should obviously reduce diffusivity in a concentration-dependent manner. We have modelled the concentration dependence of the diffusivity using a simple power law to keep the number of parameters to no more than three. Since, in thermophoresis, concentration is a function of temperature, this process ultimately introduces a specific form of temperature dependence into the diffusion currents and the $S_T$ resulting thereof.

The concentration dependence of diffusivity through the non-Fickian diffusion current influences the Soret coefficient in a very significant manner. Although the parameter $b$ ultimately gets absorbed in the normalisation of the concentration and is thereby not apparent in the expression of the Soret coefficient, the parameter $\beta$ plays a crucial role. Changing the value of the parameter $\beta$ from less than unity to greater than unity reverses the Soret coefficient curve, making a theromophobic-to-thermophilic transition a thermophilic-to-thermophobic one over the same temperature range. It is important to note that the test cases we have considered in this paper are all thermophilic-to thermophobic type transitions, and therefore the $\beta$ is greater than unity. Moreover, experimentally measured Soret coefficients can vary by at least a couple of orders of magnitude for other colloids. Since the factor of $1/(1-\beta)$ uniformly scales the Soret coefficient at all temperatures, e.g., as is evident from eq.(17), the concentration dependence of diffusivity with the closeness of $\beta$ to unity can actually capture such a wider range of magnitude. Any other required adjustments to the curvature and its vertical position can be achieved by appropriately adjusting the parameters $c$ and $a$, respectively.

On the other hand, we have also taken into consideration an explicit temperature dependence of diffusivity via the temperature dependence of the dynamic viscosity $\eta(T)$. Here, keeping in mind that this temperature dependence of the deviation of the dynamic viscosity from that of pure water can occur based on the interaction of the particle's surface with the nearby liquid layers, we have considered this dependence to be proportional to the local thermal expansion coefficient of water. We had to fix it this way, considering that if any solvent observable couples to the activity on the particle surface, that should, in general, be the density of the solvent. And these considerations worked quite well at matching the empirical profile of $S^\prime_T$ and the actual experimental data for the test cases we have considered. 

We have considered these test cases, recognising that, within the scope of the present model, strong electrostatic interactions are not considered. But, what we want to achieve at present is to produce observed profiles of $S^\prime_T$ using not more than three parameters, as the same number of parameters is considered in the empirical relation eq.\eqref{eq:16}.  At high salt concentration, multiple other forces come into existence, which can substantially modify the predominant interactions between the colloidal particle and water (solvent) considered in this model. Note that the well-known empirical temperature dependence of density and viscosity of water plays an extremely important role in the present model in diffusive and solvation force modelling. This reality can change in the presence of strong salt solutions for several reasons, including the additional field generated by the Seebeck effect, the shrinking of the hydration layer, and electrostatic screening. Moreover, it is important to note that the electrostatic double layer (EDL) formed around charged colloids can substantially influence the concentration dependence of diffusivity, which arises from hydrodynamic effects near interfaces. Exactly how the EDL will modify the concentration dependence of diffusivity is difficult to assess without detailed investigations.

It is important to note that there may be very different parameter regimes in $a, c, \beta$ in which a similar curve-fitting could be obtained for $S_T$. However, only the current parameter range used to fit the curves makes physical sense for the following reasons. (a) Since the parameter $a$ provides a vertical shift to the $S_T$ curve according to our model, it cannot be of any other order than $O$($10^{-2}$) for the test cases. (b) The magnitude of the parameter $c$ should be less than unity in order to avoid having the correction to the $\eta_w(T)$ of the same order as it. (c) Since the parameter $\beta$ flips the $S_T$ curve, it is $10>\beta >1$ to keep the order of magnitude of $S_T$ in the experimental regime besides getting the flip. We have checked that, corresponding to a relatively higher value of the $\beta$, there are no values of the parameters $a$ and $c$ in the permissible limit, and, moreover, the $C^{*}_{sat}(T)$ for Lysozyme is getting a poor fit.

We would like to conclude by noting that when non-Fickian diffusion currents may dominate transport, as during the thermophilic-thermophobic transition, neglecting them will omit an important part of the process's transport. As we have already noted in the introduction, even the theory of thermodiffusion effectively takes into account the non-Fickian diffusion current. In a broad sense, existing physical models of thermophoresis treat diffusion solely as a relaxation process based on Fickian diffusion, which slowly dissipates concentration inhomogeneities. We believe this is an incomplete physical picture, particularly given the success of the existing theory of thermodiffusion. There necessarily exist non-Fickian diffusion currents in the presence of temperature and concentration inhomogeneities; these currents can not only drive concentration instabilities but also, through their coupling to solvation forces, involve quite complex physics at microscopic scales. The scope of our model for thermophoresis and thermodiffusion will depend on future analyses along this line.

\begin{acknowledgments}
Mayank gratefully acknowledges financial support from the CSIR-SRF-Direct fellowship (Grant No. 09/0936(18027)/2024-EMR-I). Mayank dedicates this work to the loving memory of his grandmother.
\end{acknowledgments}

{\bf Author Contribution:} Mayank Sharma has contributed to computations, model building, result analysis, literature survey, preparation of all the graphs for the manuscript, and manuscript writing. Angad Singh has contributed to computations, model building, result analysis, and literature survey. A. Bhattacharyay has conceptualised the research, guided the research, and written the manuscript.

%
\newpage

\section*{Appendix}

The model parameters $(a, \beta, c)$ are determined by fitting the theoretical (model) prediction of the Soret coefficient to the experimental data using least square minimisation. For a given set of parameters, the model yields a temperature-dependent Soret coefficient $S_T^{\text{model}}(T; a,\beta,c)$, which is compared against the experimentally measured values $S_T^{\text{exp}}(T)$.

We define a least-squares error function
\begin{equation*}
\mathcal{E}(a,\beta,c) = \sum_{i} \left[ S_T^{\text{model}}(T_i; a,\beta,c) - S_T^{\text{exp}}(T_i) \right]^2,
\end{equation*}
where the sum runs over all experimental temperature points $T_i$. The optimal parameter set $(a^{*}, \beta^{*}, c^{*})$ is obtained by minimizing $\mathcal{E}$ are displayed in table~\ref{tab:ST_compare} in main text, where they are denoted by $(a,\beta, c)$ respectively. 

To explore the structure of the parameter space and assess parameter sensitivity, we display the error landscape below by showing pairs of parameters while keeping the third parameter fixed at its optimal value. A contour plot of the logarithm of the error function, $\log\mathcal{E}$, is used for visualisation purposes.

\renewcommand{\thefigure}{A1}
\begin{figure}[h]
\centering
\begin{tikzpicture}
\node[anchor=south west, inner sep=0] (img) at (0,0)
    {\includegraphics[width=\textwidth]{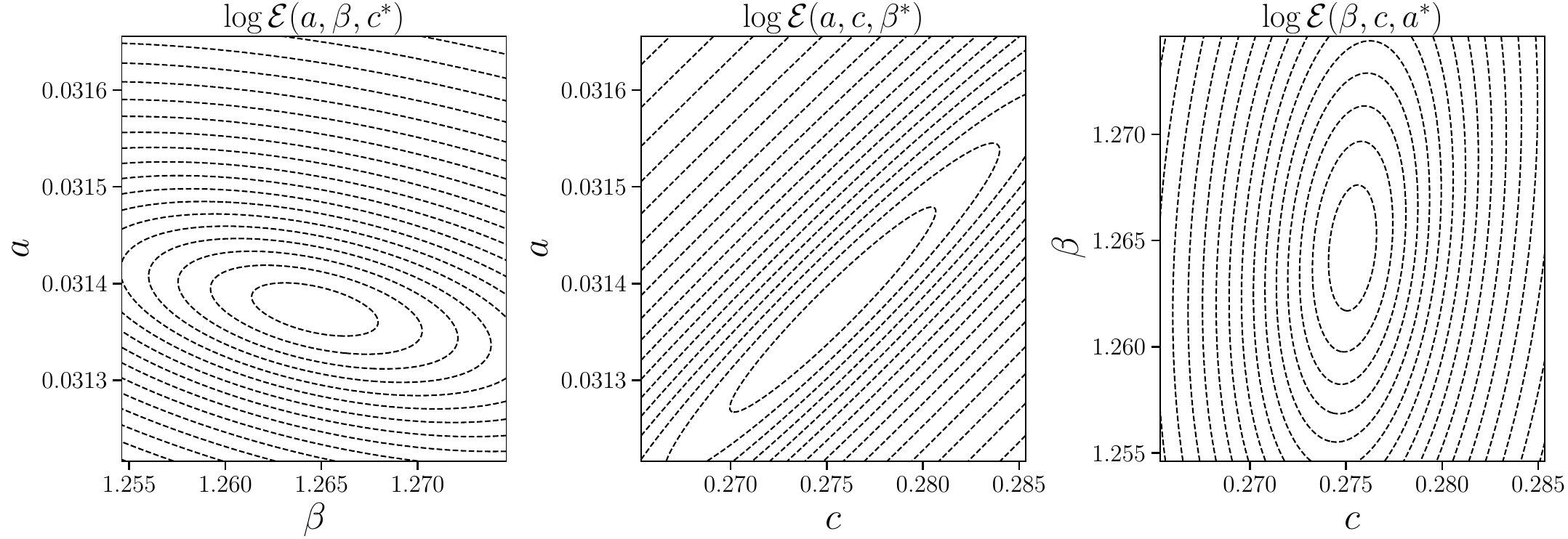}};
\begin{scope}[x={(img.south east)}, y={(img.north west)}]
    \node at (0.21,-0.12) {\large\textbf{(a)}};
    \node at (0.55,-0.12) {\large\textbf{(b)}};
    \node at (0.89,-0.12) {\large\textbf{(c)}};
\end{scope}
\end{tikzpicture}
\caption{ Contour plot of $\log \mathcal{E}$ showing the error landscape in parameter space for Lysozyme protein.}
\label{error_lysozyme}
\end{figure}

\renewcommand{\thefigure}{A2}
\begin{figure}[h]
\centering
\begin{tikzpicture}
\node[anchor=south west, inner sep=0] (img) at (0,0)
    {\includegraphics[width=\textwidth]{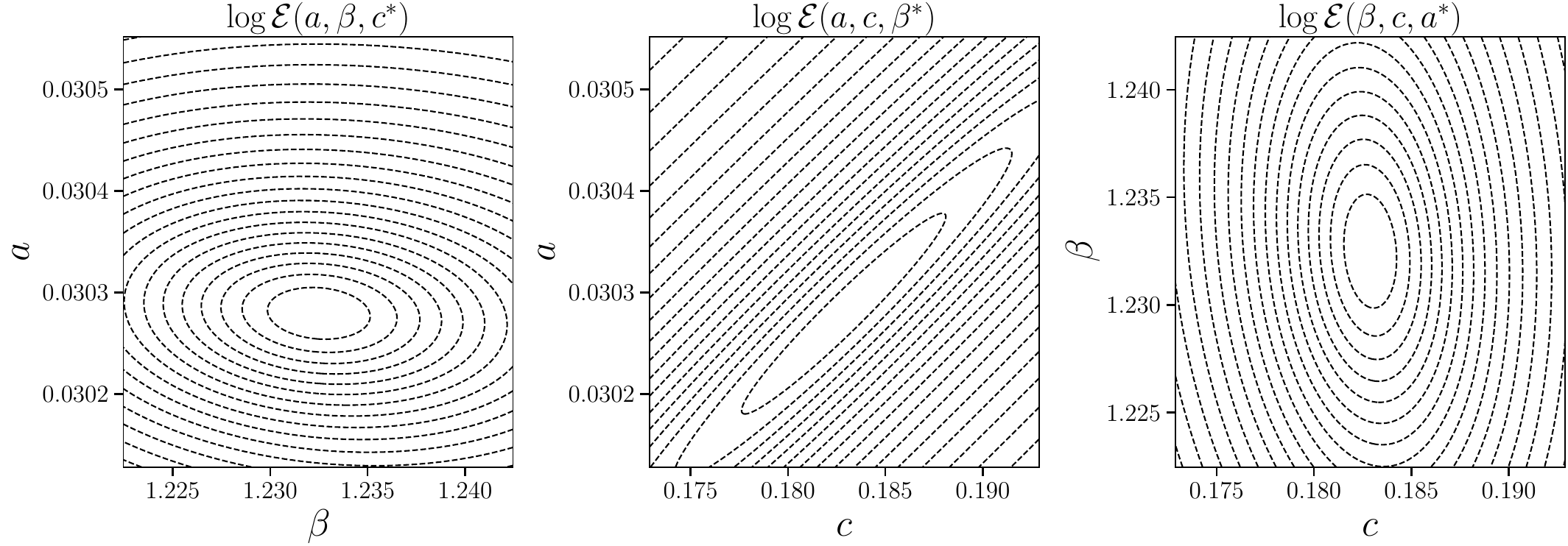}};
\begin{scope}[x={(img.south east)}, y={(img.north west)}]
    \node at (0.21,-0.12) {\large\textbf{(a)}};
    \node at (0.55,-0.12) {\large\textbf{(b)}};
    \node at (0.89,-0.12) {\large\textbf{(c)}};
\end{scope}
\end{tikzpicture}
\caption{Contour plot of $\log \mathcal{E}$ showing the error landscape in parameter space for BLGA protein. }
\label{error_blga}
\end{figure}

\renewcommand{\thefigure}{A3}
\begin{figure}[h]
\centering
\begin{tikzpicture}
\node[anchor=south west, inner sep=0] (img) at (0,0)
    {\includegraphics[width=\textwidth]{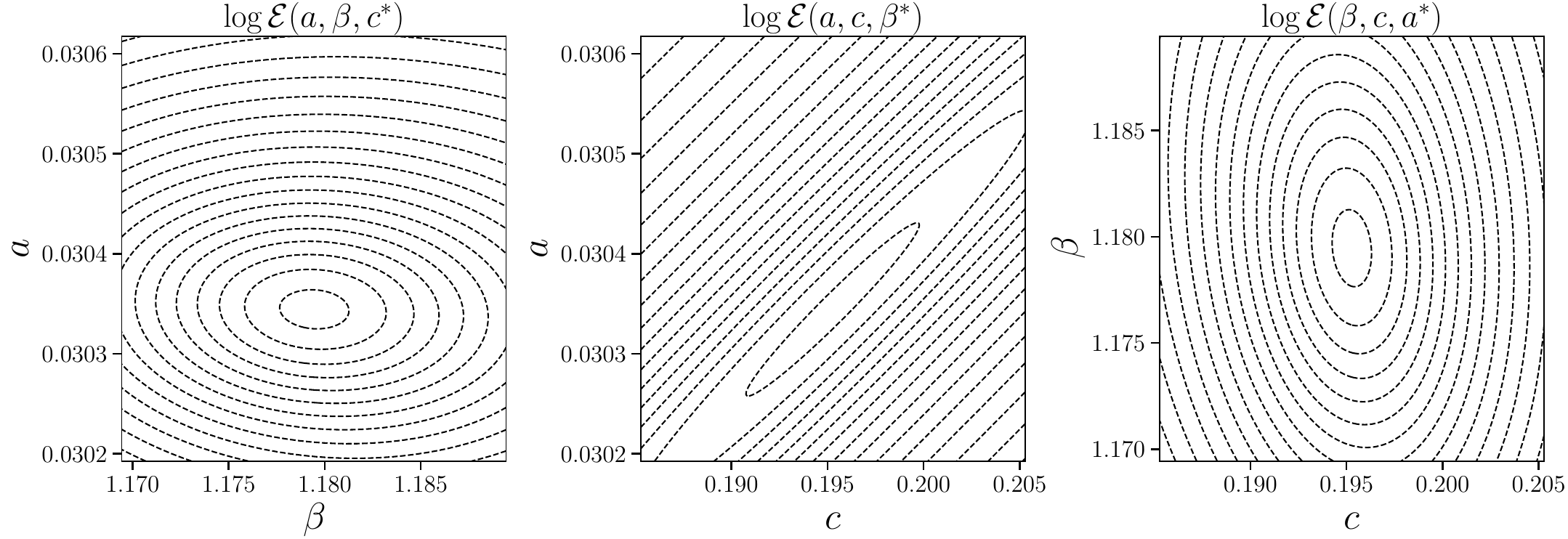}};
\begin{scope}[x={(img.south east)}, y={(img.north west)}]
    \node at (0.21,-0.12) {\large\textbf{(a)}};
    \node at (0.55,-0.12) {\large\textbf{(b)}};
    \node at (0.89,-0.12) {\large\textbf{(c)}};
\end{scope}
\end{tikzpicture}
\caption{Contour plot of $\log \mathcal{E}$ showing the error landscape in parameter space for Poly-L-Lysine protein. }
\label{error_poly_l_lysine}
\end{figure}

 \end{document}